\documentclass[%
 reprint,
superscriptaddress,
 amsmath,amssymb,
 aps,
pra,
]{revtex4-1}

\usepackage{amsthm,amssymb}
\usepackage{amsmath}
\usepackage{graphicx}
\usepackage{dcolumn}
\usepackage{bm}
\usepackage{multirow} 
\usepackage{mathdots}
\usepackage{enumerate}
\usepackage{algorithm}
\usepackage{algorithmicx}
\usepackage{algpseudocode}
\usepackage{textcomp}

\theoremstyle{remark}
\newtheorem{definition}{\indent Definition}
\newtheorem{lemma}{\indent Lemma}

\begin{document}

\makeatletter
\newcommand{\ud}{\mathrm{d}}
\newcommand{\rmnum}[1]{\romannumeral #1}
\newcommand{\polylog}{\mathrm{polylog}}
\newcommand{\ket}[1]{|{#1}\rangle}
\newcommand{\bra}[1]{\langle{#1}|}
\newcommand{\inn}[2]{\langle{#1}|#2\rangle}
\newcommand{\Rmnum}[1]{\expandafter\@slowromancap\romannumeral #1@}
\newcommand{\udots}{\mathinner{\mskip1mu\raise1pt\vbox{\kern7pt\hbox{.}}
        \mskip2mu\raise4pt\hbox{.}\mskip2mu\raise7pt\hbox{.}\mskip1mu}}
\makeatother

\preprint{APS/123-QED}

\title{Quantum algorithm for Neighborhood Preserving Embedding}

\author{Shi-Jie Pan}
\affiliation{State Key Laboratory of Networking and Switching Technology, Beijing University of Posts and Telecommunications, Beijing, 100876, China}
\affiliation{State Key Laboratory of Cryptology, P.O. Box 5159, Beijing, 100878, China}
\author{Lin-Chun Wan}
\affiliation{State Key Laboratory of Networking and Switching Technology, Beijing University of Posts and Telecommunications, Beijing, 100876, China}
\author{Hai-Ling Liu}
\affiliation{State Key Laboratory of Networking and Switching Technology, Beijing University of Posts and Telecommunications, Beijing, 100876, China}
\author{Yu-Sen Wu}
\affiliation{State Key Laboratory of Networking and Switching Technology, Beijing University of Posts and Telecommunications, Beijing, 100876, China}
\author{Su-Juan Qin}
\affiliation{State Key Laboratory of Networking and Switching Technology, Beijing University of Posts and Telecommunications, Beijing, 100876, China}
\author{Qiao-Yan Wen}
\affiliation{State Key Laboratory of Networking and Switching Technology, Beijing University of Posts and Telecommunications, Beijing, 100876, China}
\author{Fei Gao}
\email{gaof@bupt.edu.cn}
\affiliation{State Key Laboratory of Networking and Switching Technology, Beijing University of Posts and Telecommunications, Beijing, 100876, China}

\date{\today}

\begin{abstract}
Neighborhood Preserving Embedding (NPE) is an important linear dimensionality reduction technique that aims at preserving the local manifold structure. NPE contains three steps, i.e., finding the nearest neighbors of each data point, constructing the weight matrix, and obtaining the transformation matrix. Liang et al. proposed a variational quantum algorithm (VQA) for NPE [Phys. Rev. A 101, 032323 (2020)]. The algorithm consists of three quantum sub-algorithms, corresponding to the three steps of NPE, and was expected to have an exponential speedup on the dimensionality $n$.
However, the algorithm has two disadvantages: (1) It is incomplete in the sense that the input of the third sub-algorithm cannot be obtained by the second sub-algorithm. (2) Its complexity cannot be rigorously analyzed because the third sub-algorithm in it is a VQA. In this paper, we propose a complete quantum algorithm for NPE, in which we redesign the three sub-algorithms and give a rigorous complexity analysis. It is shown that our algorithm can achieve a polynomial speedup on the number of data points $m$ and an exponential speedup on the dimensionality $n$ under certain conditions over the classical NPE algorithm, and achieve significant speedup compared to Liang et al.'s algorithm even without considering the complexity of the VQA.
\end{abstract}

\pacs{Valid PACS appear here}
\maketitle

\section{Introduction}
Quantum computing theoretically demonstrates its computational advantages in solving certain problems compared with classical computing, such as the problem of factoring integers \cite{PS1994}, unstructured data searching problem \cite{GL1996} and matrix computation problems \cite{HHL,Wan2018}. In recent years, quantum machine learning has received widespread attention as a method that successfully combines classical machine learning with quantum physics. An important direction of quantum machine learning is to design quantum algorithms to accelerate classical machine learning, including data classification \cite{LMR, RML, CD}, linear regression \cite{WBL,SSP, W, YGW, YGLHRW}, association rules mining \cite{YGWW} and anomaly detection \cite{LR}.

Dimensionality Reduction (DR) is an important part of machine learning, which aims to reduce the dimensionality of the training data set while preserving the structure information of the data points as well as possible. The DR algorithm often serves as a preprocessing step in data mining and machine learning to reduce the time complexity of the algorithm and avoid a problem called \emph{curse of dimensionality} \cite{Bpattern}. Generally, The DR algorithms can be classified into two categories: the linear one and the nonlinear one. The most widely used linear DR algorithms include Principal Component Analysis (PCA) \cite{HHPCA}, Linear Discriminant Analysis (LDA) \cite{Fisher1936} and Neighborhood Preserving Embedding (NPE) \cite{he2005neighborhood}, while the typical nonlinear DR algorithm is Locally Linear Embedding (LLE) \cite{roweis2000nonlinear}.
Here, we focus on NPE which can be regarded as the linear approximation of LLE. Unlike PCA that tries to preserve the global Euclidean structure, NPE aims at preserving the local manifold structure. Furthermore, NPE has a closed-form solution. Similar to other DR algorithms, NPE requires a large amount of computational resources in the big-data scenario because of its high complexity.

In recent years, some researchers successfully combined DR algorithms with quantum techniques and obtained various degrees of speedups. Lloyd et al. proposed a quantum PCA algorithm to reveal the large eigenvectors in quantum form of an unknown low-rank density matrix, which achieves an exponential speedup on the dimension of the training data \cite{Lloyd_2014}. Latter, Yu et al. proposed a quantum algorithm that compresses training data based on PCA \cite{YGLW}, and achieves an exponential speedup on the dimension over the classical algorithm. Cong et al. proposed a quantum LDA algorithm for classification with exponential speedups on the scales of the training data over the classical algorithm \cite{CD}.
Besides, there are some other quantum DR algorithms, including quantum A-optimal projection \cite{DYXL,Pan2020}, quantum kernel PCA \cite{Li_2020} and quantum spectral regression \cite{MYX}.

For NPE, Liang et al. proposed a Variational Quantum Algorithm (VQA)\cite{Liang2020}, called VQNPE. NPE contains three steps, i.e., finding the nearest neighbors of each data point, constructing the weight matrix, and obtaining the transformation matrix $A$. VQNPE includes three sub-algorithms, corresponding to the three steps of NPE. However, VQNPE has two drawbacks: (1) The algorithm is incomplete. As the authors pointed out, it is not known how to obtain the input of the third sub-algorithm from the output of the second one. (2) It lacks a provable quantum advantage. Since the advantage of VQA has not been proved rigorously yet (generally, we say that VQA has potential advantage \cite{MARS,LWWP}), it is hard to exam the speedups of Liang et al.'s algorithm.

In this paper, we propose a complete quantum NPE algorithm with rigorous complexity analysis. Our quantum algorithm also consists of three quantum sub-algorithms, corresponding to the three steps of the classical NPE.
The first one is finding the neighbors of each data point by quantum amplitude estimation and quantum amplitude amplification. By storing the information
of neighbors in a data structure of QRAM \cite{IA,WZP}, we obtain two
oracles. With these oracles, the second one reveals the classical information of the weight matrix $W$ column by column by quantum matrix inversion technique. In the third one, we use a quantum version of the Spectral Regression (SR) method (a modification of \cite{MYX}) to get the transformation matrix $A$. Specifically, we obtain the $d$ ($d$ is the dimension of the low dimensional space) bottom nonzero eigenvectors of the matrix $M=(I-W)^T(I-W)$ at first, and then perform several times of the quantum ridge regression algorithm to obtain $A$.
As a conclusion, under certain conditions, our algorithm has a polynomial speedup on the number of data points $m$ and exponential speedup on the dimension of the data points $n$ over the classical NPE algorithm, and has a significant speedup compared with even the first two sub-algorithms of VQNPE.

The rest of this paper is organized as follows. In Sec. \Rmnum{2}, we review the classical NPE algorithm. In Sec.  \Rmnum{3}, we propose our quantum NPE algorithm and analyze the complexity. Specifically, in Sec. \Rmnum{3} A, we propose a quantum algorithm to find the nearest neighbors of each data point and analyze the complexity. In  Sec. \Rmnum{3} B, we propose a quantum algorithm to obtain the information of the weight matrix $W$ and analyze the complexity. The quantum algorithm for computing the transformation matrix $A$ is proposed in Sec. \Rmnum{3} C, together with the complexity analysis. The algorithm procedures and the complexity is concluded in \Rmnum{3} D, along with a comparison with VQNPE. The conclusion is given in Sec. \Rmnum{4}.

\section{Review of the classical NPE}
\label{sec:2}
In this section, we briefly review the classical NPE \cite{he2005neighborhood,roweis2000nonlinear,chenlocally}.

Suppose $X=(\textbf{x}_0, \textbf{x}_1,..., \textbf{x}_{m-1})^T$ is a data matrix with dimension $m \times n$, where $n$ is the dimension of $\textbf{x}_{i}$ and $m$ is the number of data points. The objective of NPE is to find a matrix $A$ (called transformation matrix) embedding the data matrix into a low-dimensional space (assume the embedding results is $\mathbf{y}_0,\mathbf{y}_1,...,\mathbf{y}_{m-1}$, $\mathbf{y}_{i}\in \mathbb{R}^d$ and $d\ll n$, we have $\mathbf{y}_i=A^T \mathbf{x}_i$, $A\in \mathbb{R}^{n\times d}$) that the linear relation between each data point and its nearest neighbors is best preserved.
Specifically, suppose the nearest neighbors of $\textbf{x}_i$ are  $\textbf{x}_j, \textbf{x}_k$ and $\textbf{x}_l$, then $\textbf{x}_i$ can be reconstructed (or approximately reconstructed) by linear combination of $\textbf{x}_j, \textbf{x}_k, \textbf{x}_l$, that is,
\begin{eqnarray}\label{eq:defW}
            \textbf{x}_i=W_{ij}\textbf{x}_j+W_{ik}\textbf{x}_k+W_{il}\textbf{x}_l,
\end{eqnarray}
where $W_{ij},W_{ik}$ and $W_{il}$ are weights that summarize the contribution of $\textbf{x}_j,\textbf{x}_k$ and $\textbf{x}_l$ to the reconstruction of $\textbf{x}_i$. NPE trys to preserve the linear relations in Eq. (\ref{eq:defW}) in the low-dimensional embedding.

NPE consists of the following three steps.

Step 1: Find the nearest neighbors of each data point. There are two most common techniques to find the nearest neighbors. One is $k$-Nearest Neighbors algorithm (kNN) with a fixed $k$, and the other is choosing neighbors within a ball of fixed radius $r$ based on Euclidean distance for each data point.

Step 2: Construct the weight matrix $W\in \mathbb{R}^{m\times m}$, where the $(i+1)$th row and $(j+1)$th column element is $W_{ij}$. Suppose the set of the nearest neighbors of the data point $\textbf{x}_i$ is denoted as $Q_i$, then the construction of $W$ is to optimize the following objective function:
    \begin{eqnarray}\label{eq:optimizeW}
    \begin{split}
            &\mathop{\min}_W \sum_{i=0}^{m-1} \left\|\textbf{x}_i-\sum_{j \in Q_i} W_{ij}\textbf{x}_j\right\|^2,\\
            &s.t. \quad \quad \quad \sum_{j \in Q_i} W_{ij}=1.
    \end{split}
    \end{eqnarray}
Note that the data point $\textbf{x}_i$ is only reconstructed by its nearest neighbors, i.e, the elements in $Q_i$. If $\textbf{x}_j \notin Q_i$, we set $W_{ij}=0$. We should mention that $\|\bullet\|$ is the $L_2$ norm of a vector or the spectral norm of a matrix in this paper. The above optimization problem has a closed form solution. Let $C^{(i)}$ denote an $m \times m$ matrix related to $\textbf{x}_i$, called neighborhood
correlation matrix, where
    \begin{eqnarray}\label{eq:constructC}
    \begin{split}
        C^{(i)}_{jk}=
    \begin{cases}
                (\textbf{x}_i -\textbf{x}_j)^T (\textbf{x}_i-\textbf{x}_k), &j,k \in Q_i; \\
                0, & \mbox{otherwise}.
             \end{cases}
    \end{split}
    \end{eqnarray}
Assume the number of elements of $Q_i$ is $k^{(i)}$ and $k^{(i)} \ll m$, then $C^{(i)}$ are low-rank matrices for $i\in \{0,1,...,m-1\}$. Let
$C^{(i)}=\sum^{k^{(i)}-1}_{j=0} \lambda_j^{(i)} \textbf{u}^{(i)}_j \textbf{u}^{(i)\dag}_j$, then the pseudo inverse of $C^{(i)}$ is $[C^{(i)}]^{-1}=\sum_{\lambda_j^{(i)}\neq 0} \frac{1}{\lambda_j^{(i)}} \textbf{u}^{(i)}_j \textbf{u}^{(i)\dag}_j$. Let $W_i$ denotes the $(i+1)$th row of matrix $W$, then the solution of the objective function is
    \begin{eqnarray}\label{eq:solutionW}
    \begin{split}
            W_i=\frac{[C^{(i)}]^{-1} \textbf{1}}{\textbf{1}^T[ C^{(i)}]^{-1} \textbf{1}},
    \end{split}
    \end{eqnarray}
where $\textbf{1}=(1,1,...,1)^T$.

Step 3: Compute the transformation matrix $A$. To best preserve the linear relations in the low-dimensional space, the optimization problem is designed as follows:
\begin{eqnarray}\label{eq:W}
    \begin{split}
            \mathop{\min}_{A} &\sum_{i=0}^{m-1} \left\|\textbf{y}_i-\sum_{j \in Q_i} W_{ij}\textbf{y}_j\right\|^2,\\
            s.t.  \quad  &\sum_{i=0}^{m-1} \mathbf{y}_i=\mathbf{0}, \quad
            \frac{1}{m}\sum_{i=0}^m \mathbf{y}_i \mathbf{y}_i^T=I,\\
            &\quad \mathbf{y}_i=A^T \mathbf{x}_i,
    \end{split}
    \end{eqnarray}
where $\textbf{y}_0, \textbf{y}_1,..., \textbf{y}_{m-1}$ are the low-dimensional embeddings. The optimization problem can be minimized by solving the following generalized eigen-problem:
\begin{eqnarray}\label{eq:optimizeM}
    \begin{split}
        X^T MX \mathbf{a}=\lambda X^T X \mathbf{a},
    \end{split}
    \end{eqnarray}
where $M$ is a sparse matrix that equates $(I-W)^T(I-W)$. Then the bottom $d$ nonzero eigenvectors $\textbf{a}_0, \textbf{a}_1,..., \textbf{a}_{d-1}$ of the above eigen-problem with corresponding eigenvalues $0<\lambda_0 \le \lambda_1 \le ... \le \lambda_{d-1}$ yield $A=(\textbf{a}_0, \textbf{a}_1,..., \textbf{a}_{d-1})$.

There are many different methods to solve the eigenvalue problem in Eq. (\ref{eq:optimizeM}). Here we use the method mentioned in \cite{CHHSR,CHHSR1}, called Spectral Regression (SR) method. The eigenvalue problem in Eq. (\ref{eq:optimizeM}) can be solved by two steps according to the SR method. (1) Solve the following eigen-problem to get the bottom non-zero eigenvectors $\mathbf{z}_0,\mathbf{z}_1,...,\mathbf{z}_{d-1}$:
\begin{eqnarray}
    \begin{split}
    \label{mz}
        M\mathbf{z}=\lambda \mathbf{z}.
    \end{split}
    \end{eqnarray}
(2) Find $\mathbf{a}_{i}$ that satisfies
\begin{eqnarray}
    \begin{split}
       \mathbf{a}_{i} &= \arg \min_\mathbf{a} \left(\sum_{j=1}^{m} (\mathbf{a}^T \mathbf{x}_j- \mathbf{z}_{ij})^2
                        + \alpha \|\mathbf{a}\|^2\right)\\
                      &=\left(X^T X +\alpha I\right)^{-1}X^T \mathbf{z}_i,
    \end{split}
    \end{eqnarray}
where $\mathbf{z}_{ij}$ is the $j$ element of $\mathbf{z}_{i}$, $\alpha \ge 0$ is a constant to control the penalty of the norm of $\mathbf{a}$.

As a conclusion, the detailed procedures of NPE are given in Algorithm \ref{Algorithm1}.
\begin{algorithm}[H]
\caption{The procedure of NPE}
\label{Algorithm1}
\begin{algorithmic}[1]
\Require
The data set $X=(\textbf{x}_0, \textbf{x}_1,..., \textbf{x}_{m-1})^T$;
\Ensure
The transformation matrix $A=(\mathbf{a}_0,\mathbf{a}_1,...,\mathbf{a}_{d-1})$;
\State Find the set of nearest neighbors $Q_i$ of each data point $i$;
\State Construct $C^{(i)}$ by Eq. (\ref{eq:constructC}) for $i=0,1,...,m-1$;
\State Obtain $W$ by Eq. (\ref{eq:solutionW});
\State Decompose the matrix $M=(I-W)^T (I-W)$ to get the bottom $d$ nonzero eigenvectors $\mathbf{z}_1,\mathbf{z}_2,...,\mathbf{z}_{d}$;
\State Compute $\mathbf{a}_i=\left(X^T X +\alpha I\right)^{-1}X^T \mathbf{z}_i$ for $i=0,1,...,d-1$;
\Return $A$;
\end{algorithmic}
\end{algorithm}

As for the time complexity of NPE algorithm, the procedure to find the $k$ nearest neighbors of each data point has complexity $O(mn\log_{2} k\log_{2} m)$ by using BallTree \cite{scikit-learn}. The complexity to construct the weight matrix W is $O(mnk^3)$ (generally, $k\ll m$). And the procedure to get the transformation matrix $A$ has complexity $O(dm^2)$. Thus the overall complexity of NPE algorithm is $O(mnk^3 +dm^2)$.
\section{Quantum algorithm for NPE}

In this section, we introduce our quantum algorithm for NPE \cite{chenlocally,he2005neighborhood}. The quantum algorithm can be divided into three parts, corresponding to the three parts of the classical algorithm. We give a quantum algorithm to find the nearest neighbors algorithm in Sec. \ref{subsecA}, a quantum algorithm to construct the weight matrix $W$ in Sec. \ref{subsecB} and a quantum algorithm to compute the transformation matrix $A$ in Sec. \ref{subsecC}. In Sec. \ref{subsecD}, we conclude the complexity of our quantum algorithm and make a comparison with VQNPE.

\subsection{Quantum algorithm to find the nearest neighbors}
\label{subsecA}



Assume that the data matrix $X=(\textbf{x}_0, \textbf{x}_1,..., \textbf{x}_{m-1})^T$ is stored in a structured QRAM which allows the following mappings to be performed in time $O[\polylog(mn)]$ \cite{IA,WZP}:
\begin{eqnarray}\label{eq:accessX}
    \begin{split}
        &O_X: \ket{i}\ket{j}\ket{0}\rightarrow \ket{i}\ket{j}\ket{X_{ij}},\\
        &U_X: \ket{i}\ket{0}\rightarrow \frac{1}{\|X_{i \cdot}\|}\sum_{j=1}^n X_{ij}\ket{i,j}= \ket{i}\ket{\mathbf{x}_i},\\
        &V_X: \ket{0}\ket{j} \rightarrow \frac{1}{\|X\|_F}\sum_{i=1}^m \|X_{i \cdot}\|\ket{i,j},
\end{split}
\end{eqnarray}
where $X_{i \cdot}$ is the $i$th row of $X$, i.e., $\textbf{x}_i$.


In our quantum algorithm, we choose neighbors within a ball of fixed radius $r$ based on Euclidean distance for each data point (our algorithm can also be generalized to kNN to get a similar speedup). The selection of $r$ is important for the performance of this type of algorithms, but how to choose a suitable $r$ is outside the scope of our discussion. Here we assume that $r$ is constant that given in advance.
Let $k^{(i)}$ denote the number of nearest neighbors of $\textbf{x}_i$, the objective of our quantum nearest neighbors algorithm is to output the index $j$ of all $\mathbf{x}_j\in Q_i$, where $Q_i=\{\mathbf{x}_j \big|\|\mathbf{x}_i-\mathbf{x}_j\|^2 \le r^2, j\neq i, j\in {0,1,...m-1}\}$ for $i\in \{0,1,...,m-1\}$, $|Q_i|=k^{(i)}$.

\subsubsection{Algorithm details}
\label{al1}

We adopt the quantum amplitude estimation \cite{BHMT} and amplitude amplification \cite{GL1996,BHMT} to get the neighbors of $\mathbf{x}_i$. The algorithm can be decomposed into the following two stages:
\begin{enumerate}
    \item Prepare the following quantum state by quantum amplitude estimation \cite{BHMT},
    \begin{eqnarray}
    \begin{split}
        \ket{\phi}=\frac{1}{m}\sum_{i,j=0}^{m-1} \ket{i}\ket{j} \ket{\sqrt{K/m^2}},
    \end{split}
    \end{eqnarray}
    where $K$ is the number of the pairs of points that satisfy $\|\mathbf{x}_i-\mathbf{x}_j\| \le r$.
    \item Prepare quantum state $\sqrt{p} \ket{\psi} + \sqrt{1-p} \ket{\psi^\bot}$, $p>1/2$ by quantum amplitude amplification \cite{GL1996,BHMT}, where
    \begin{eqnarray}
    \begin{split}
        \ket{\psi}=\frac{1}{\sqrt{K}}\sum_{i=0}^{m-1} \ket{i}\sum_{\mathbf{x}_j \in Q_i}\ket{j},
    \end{split}
    \end{eqnarray}
    $\ket{\psi^\bot}$ is a quantum state that is orthogonal to to $\ket{\psi}$. Then by measuring the state in computational basis for several times, we could obtain the index $j$ of the neighbors of $\mathbf{x}_i$ for $i=0,1,...,m-1$.
\end{enumerate}

Here we list two lemmas that will be used in our algorithm:
\begin{lemma} (\cite{neu19})
\label{Le:distance}
Assume that $U: U\ket{i}\ket{0}=\ket{i}\ket{\mathbf{v}_i}$ and $V: V\ket{j}\ket{0}=\ket{j}\ket{\mathbf{c}_j}$ can be performed in time $T$, and the norms of the vectors $\mathbf{v}_i$ and $\mathbf{c}_j$ are known. Let $d^2(\mathbf{v}_i,\mathbf{c}_j)=\|\mathbf{v}_i-\mathbf{c}_j\|^2$, then a quantum algorithm can compute
    \begin{eqnarray}\label{distance}
            \ket{i}\ket{j}\ket{0}\mapsto \ket{i}\ket{j}\ket{d^2(\mathbf{v}_i,\mathbf{c}_j)},
    \end{eqnarray}
with probability at least $1-2\delta$ for any $\delta$ with complexity $O(\frac{\|\mathbf{v}_i\|\|\mathbf{c}_j\|T \log_{2}(1/\delta)}{\epsilon})$, where $\epsilon$ is the error of $d^2(\mathbf{v}_i,\mathbf{c}_j)$.
\end{lemma}

We now detail the stage 1. We first prepare the state $\frac{1}{m}\sum_{i,j=0}^{m-1} \ket{i}\ket{j}$.
According to \emph{Lemma} \ref{Le:distance}, we can obtain the state $\frac{1}{m}\sum_{i,j=0}^{m-1} \ket{i}\ket{j} \ket{\|\mathbf{x}_i-\mathbf{x}_j\|^2}$ by a unitary (denotes as $U_1$) with complexity $O\left[\frac{(\max_i \|\mathbf{x}_i\|)^2 T \log_{2}(1/\delta)}{\epsilon_1}\right]$, where $T=O[\polylog(mn)]$ is the complexity of the mappings in Eq. (\ref{eq:accessX}), $1-2\delta$ is the successful probability and $\epsilon_1$ is the error of the value of $\|\mathbf{x}_i-\mathbf{x}_j\|^2$ stored in the third register.

Then, let $O_1$ be the unitary that transforms $\frac{1}{m}\sum_{i,j=0}^{m-1} \ket{i}\ket{j} \ket{\|\mathbf{x}_i-\mathbf{x}_j\|^2}$ to the state
    \begin{eqnarray*}\label{eq:Oracle}
    \begin{split}
        \frac{1}{m}\sum_{i=0}^{m-1}\ket{i} \left(\sum_{\mathbf{x}_j \notin Q_i}\ket{j} \ket{\|\mathbf{x}_i-\mathbf{x}_j\|^2}-
        \sum_{\mathbf{x}_j \in Q_i}\ket{j} \ket{\|\mathbf{x}_i-\mathbf{x}_j\|^2}\right),
    \end{split}
    \end{eqnarray*}
thus $O=U_1^{-1}O_1 U_1$ transforms $\frac{1}{m}\sum_{i,j=0}^{m-1} \ket{i}\ket{j}$ to
    \begin{eqnarray}\label{eq:Oracle1}
    \begin{split}
                \frac{1}{m}\sum_{i=0}^{m-1}\ket{i} \left(\sum_{\mathbf{x}_j \notin Q_i}\ket{j}-
        \sum_{\mathbf{x}_j \in Q_i}\ket{j}\right).
    \end{split}
    \end{eqnarray}
We can perform quantum amplitude estimation \cite{BHMT} on $\frac{1}{m}\sum_{i,j=0}^{m-1} \ket{i}\ket{j}$ with oracle $O$ and Grover operator $G$, where
    \begin{eqnarray*}
    \begin{split}
        G=H^{\otimes 2\log_{2} m}(2\ket{0}\bra{0}^{2\log_{2} m}-I_{m^2 \times m^2})H^{\otimes 2\log_{2} m}O.
    \end{split}
    \end{eqnarray*}
The output state of quantum amplitude estimation is $\ket{\phi}=\frac{1}{m}\sum_{i,j=0}^{m-1} \ket{i}\ket{j} \ket{\sqrt{K/m^2}}$.

For the stage 2, based on the output of stage 1, we prepare the state $\ket{\phi_1}=\frac{1}{m}\sum_{i=0}^{m-1} \ket{i} \sum_{j=0}^{m-1} \ket{j} \ket{t}$, where $t=\lceil \frac{\pi}{4}\sqrt{\frac{m^2}{K}} \rceil$. Then we apply quantum amplitude amplification \cite{GL1996,BHMT} to $\ket{\phi_1}$. Specifically, we apply $t$ times of $G$ operator, which controlled by the third register of $\ket{\phi_1}$, i.e.,
\begin{eqnarray*}
            \frac{1}{m}\ket{t}\sum_{i=0}^{m-1} \ket{i} \sum_{j=0}^{m-1} \ket{j} \rightarrow
            \frac{1}{m}\ket{t} G^{t}\sum_{i=0}^{m-1} \ket{i} \sum_{j=0}^{m-1} \ket{j}.
\end{eqnarray*}
Thus we can obtain the quantum state
\begin{eqnarray}
\label{PP}
\ket{t} \left(\sqrt{p} \ket{\psi} + \sqrt{1-p} \ket{\psi^\bot}\right),
\end{eqnarray}
where $\ket{\psi^\bot}$ is the quantum state that is orthogonal to $\ket{\psi}$.
If the estimation of $t$ is sufficiently precise, for example, within error $t/3$, we have $p> 1/2$. Finally,
by discarding the first register, we could get $\sqrt{p} \ket{\psi} + \sqrt{1-p} \ket{\psi^\bot}$.

\subsubsection{Complexity analysis}

To guarantee $p > 1/2$ of Eq. (\ref{PP}), the error of $t$ should be less than $t/3$. Since $t=\lceil \frac{\pi}{4}\sqrt{\frac{m^2}{K}} \rceil$, we could make the error of $K$ no more than $\frac{1}{2} K$.

In stage 1, since $O_1$ can be implemented in $O[\polylog(mn)]$ and $U_1$ has complexity $O\left[\frac{(\max_i \|\mathbf{x}_i\|)^2 T \log_{2}(1/\delta)}{\epsilon_1}\right]$, the complexity of the oracle $O=U_1^{-1}O_1 U_1$ is $T_{oracle}=O\left[\frac{(\max_i \|\mathbf{x}_i\|)^2 T \polylog(mn/\delta)}{\epsilon_1}\right]$.
The number of queries of $O$ in quantum amplitude estimation is $O(\sqrt{m^2 K}/\epsilon_K)=O(\sqrt{m^2/K})$, $\epsilon_K=\frac{1}{2} K$ is the estimate error of $K$. Assume that the number of neighbors of each point $\mathbf{x}_i$ is balanced, that is, $k^{(i)}=\Theta(k)$ for $i=0,1,...,m-1$, $k^{(i)}$ is the number of neighbors of $\mathbf{x}_i$ (we should mention that if we adopt the kNN algorithm, $k^{(i)}=k$). Thus $K=\sum_i k^{(i)}=\Theta(mk)$. Therefore, the time complexity of stage 1 is $T_{s1}^{(1)}=O\left[\frac{(\max_i \|\mathbf{x}_i\|)^2 T \sqrt m \polylog(mn/\delta)}{\epsilon_1\sqrt{k}}\right]$.

In stage 2, $t$ times of $G$ is implemented in the quantum amplitude amplification step to obtain the state in Eq. (\ref{PP}), where $t=\lceil \frac{\pi}{4}\sqrt{\frac{m^2}{K}}\rceil=O(\sqrt{\frac{m}{k}})$. Since the error of $t$ is less than $t/3$, we have $p>1/2$. We denote the complexity to get the state in Eq. (\ref{PP}) as $T_{s2}^{(1)}$.
To reveal all pairs $(i,j)$ that satisfy $\|\mathbf{x}_i-\mathbf{x}_j\| \le r$, we should measure the state $\ket{\psi}$ for $O(K\log_{2} K)$ times \cite{erdHos1961classical}.

The $\delta$ would transform to an error $
O(\sqrt{\frac{m}{k}}\delta)$ in $\ket{\psi}$, let $\delta=O(\sqrt{\frac{k\epsilon^2}{m}})$, then the final error of $\ket{\psi}$ is $\epsilon$. The $\epsilon_1$ is related to the actual data set and the choice of $r$, here we assume it to be a constant.
Let $h=\max_{i} \|\mathbf{x}_i\|$, note that $T=O[\polylog(mn)]$, the total complexity of this algorithm is
\begin{eqnarray}
\begin{split}
           T^{(1)}&=O(K \log_{2} K(T_{s1}^{(1)}+T_{s2}^{(1)}))\\
           &=O\left[h^2 m^{3/2} k^{1/2} \polylog(mn/\epsilon)\right].
\end{split}
\end{eqnarray}


\subsection{Quantum algorithm to obtain the weight matrix $W$}
\label{subsecB}

Since we have obtained the indexes of the neighbors of all the data points in the previous algorithm. These information can be represented as a matrix $B$ with $B_{ij}=1$ if $\mathbf{x}_j\in Q_i$ and $B_{ij}=0$ if $\mathbf{x}_j\notin Q_i$. To facilitate quantum access in the subsequent algorithms, we store the matrix $B$ in a data structure \cite{IA}, that allows the following two mappings
\begin{eqnarray}
\begin{split}
\label{oracleA}
        &U_B: \ket{i}\ket{0} \mapsto  \ket{i}\ket{B_i},\\
        &V_B: \ket{0}\ket{j} \mapsto \frac{1}{\sqrt{\|B\|_F}}\sum_i \|B_i\|\ket{i}\ket{j}
\end{split}
\end{eqnarray}
in complexity $O[\polylog (mn)]$, where
\begin{eqnarray}
\begin{split}
\label{ai}
        \ket{B_i}=\frac{1}{\sqrt{k^{(i)}}}\sum_{\mathbf{x}_j\in Q_i} \ket{j}.
\end{split}
\end{eqnarray}

The size of the data structure is $O[K \log_{2}^2(mn)]=O[mk \log_{2}^2(mn)]$, and the time to store the matrix $B$ in the data structure is $O[K \log_{2}^2(mn)]=O[mk \log_{2}^2(mn)]$. We should mention that the complexity to construct the data structure (that is, store $B$ in the data structure) can be neglected, since the complexity to obtain the matrix $B$ is much greater.

Let $\rho_{C^{(i)}}$ be a dense matrix that is proportional to $C^{(i)}$, that is $\rho_{C^{(i)}}\propto C^{(i)}$, for $i=0,1,2,...,m-1$. Then the weight matrix $W=(W_1,W_2,...,W_m)$ with
\begin{eqnarray}
\begin{split}
    W_i=\frac{[C^{(i)}]^{-1} \textbf{1}}{\textbf{1}^T[ C^{(i)}]^{-1} \textbf{1}}
    \propto \frac{\rho_{C^{(i)}}^{-1}\ket{B_i}}{\|\rho_{C^{(i)}}^{-1}\ket{B_i}\|}:=\ket{W_i}.
\end{split}
\end{eqnarray}
To obtain each $W_i$, We give a quantum algorithm to prepare a state which is a purification of $\rho_{C^{(i)}}$ first. Then by means of this quantum algorithm, we can perform matrix inversion of $\rho_{C^{(i)}}$ on the state $\ket{B_i}=\frac{1}{\sqrt{k^{(i)}}}\sum_{\mathbf{x}_j\in Q_i} \ket{j}$ to obtain $\ket{W_i}$. Finally, quantum state tomography is used to reveal the information of $\ket{W_i}$. Since the state $\ket{W_i}$ is sparse, the quantum tomography step is efficient.

According to Eq. (\ref{eq:constructC}), we have
\begin{eqnarray*}
    \begin{split}
        C^{(i)}=\sum_{\mathbf{x}_j, \mathbf{x}_k\in Q_i} \|\mathbf{x}_i-\mathbf{x}_j\| \|\mathbf{x}_i-\mathbf{x}_k\| \inn{\mathbf{x}_i-\mathbf{x}_j}{\mathbf{x}_i-\mathbf{x}_k}\ket{j}\bra{k},
    \end{split}
    \end{eqnarray*}
where $\ket{\mathbf{x}_i-\mathbf{x}_j}$ denotes the quantum state which is proportional to the vector $\mathbf{x}_i-\mathbf{x}_j$, i.e.,
\begin{eqnarray}
    \ket{\mathbf{x}_i-\mathbf{x}_j}= \frac{\mathbf{x}_i-\mathbf{x}_j}{\|\mathbf{x}_i-\mathbf{x}_j\|}.
\end{eqnarray}
Let
\begin{eqnarray}
            \ket{\psi^{(i)}}= \ket{i}\frac{1}{\sqrt{c^{(i)}}}\sum_{\mathbf{x}_j\in Q_i} \|\mathbf{x}_i-\mathbf{x}_j\|\ket{j}\ket{\mathbf{x}_i-\mathbf{x}_j},
    \end{eqnarray}
where $c^{(i)}=\sum_{\mathbf{x}_j\in Q_i}\|\mathbf{x}_i-\mathbf{x}_j\|^2$ is the normalized factor. $\ket{\psi^{(i)}}$ is a purification of $\rho_{C^{(i)}}$, since by taking partial trace of the first and third registers, we can obtain
     \begin{eqnarray}
     \begin{split}
            \rho_{C^{(i)}}=&\frac{1}{c^{(i)}}\sum_{\mathbf{x}_j,\mathbf{x}_k\in Q_i} \|\mathbf{x}_i-\mathbf{x}_j\|\|\mathbf{x}_i-\mathbf{x}_k\|  \\
            &\inn{\mathbf{x}_i-\mathbf{x}_j}{\mathbf{x}_i-\mathbf{x}_k} \ket{j}\bra{k}.
     \end{split}
    \end{eqnarray}

\subsubsection{Algorithm details}

The algorithm to obtain each row of the weight matrix $W$, i.e., $W_i$, can be decomposed to the following three stages:
\begin{enumerate}
    \item Prepare the quantum state $\ket{\psi^{(i)}}$.
    \item Prepare quantum state $\ket{W_i}=\rho_{C^{(i)}}^{-1}\ket{B_i}$ by quantum matrix inversion technique \cite{chakraborty2019,Lloyd_2014}.
    \item Perform quantum state tomography on the state $\ket{W_i}$ to get the information of $W_i$.
\end{enumerate}

Here we list a definition and three lemmas that will be used in our algorithm:
\begin{definition}(\cite{Yuansu})
\label{def:block}
An $(n+a)$-qubit unitary $U$ is called an $(\alpha,a,\epsilon)$ block-encoding of a matrix $A \in \mathbb{C}^{n\times n}$ if it satisfies
   \begin{eqnarray}
            \|\alpha (\bra{0}^{\otimes a} \otimes I)U (\ket{0}^{\otimes a} \otimes I)-A\| \le \epsilon,
   \end{eqnarray}
where $\alpha>0$.
\end{definition}

\begin{lemma}
\label{minus}
Given oracle $O_1,O_2,O'_1$ and $O'_2$ to access the vectors $\mathbf{x}_i,\mathbf{y}_i$ and the norm of the vectors in time $O(\polylog (mn))$, i.e.,
   \begin{eqnarray*}
            O_1 \ket{i}\ket{0} &= \ket{i} \ket{\mathbf{x}_i}, \quad
            O'_1 \ket{i}\ket{0} &= \ket{i} \ket{\|\mathbf{x}_i\|}; \\
            O_2 \ket{i}\ket{0} &= \ket{i} \ket{\mathbf{y}_i}, \quad
            O'_2 \ket{i}\ket{0} &= \ket{i} \ket{\|\mathbf{y}_i\|},
   \end{eqnarray*}
there exists a quantum algorithm converts
   \begin{eqnarray}
            \sum_{i,j=0}^{m-1} \sqrt{p_{ij}}\ket{i}\ket{j}\rightarrow\sum_{i,j=0}^{m-1} \sqrt{p_{ij}}\ket{i}\ket{j}\ket{\mathbf{x}_i-\mathbf{y}_j},
   \end{eqnarray}
with complexity $O(\frac{h}{\epsilon_0}\polylog(mn/\epsilon))$, where $h=\max_{i} \{\|\mathbf{x}_i\|,\|\mathbf{y}_i\|\}$, $\epsilon$ is the error of the output state and $\epsilon_0=\min_{i,j} \|\mathbf{x}_i-\mathbf{y}_j\|$.
\end{lemma}

Proof: see Appendix \ref{app:aa}.

\begin{lemma}(\cite{Low2019hamiltonian,Yuansu})
\label{blockencode}
Let $G$ be an $(n+s)$-qubit unitary that generate $\rho$ by tracing out the ancillary register, that is,
\begin{eqnarray*}
\begin{split}
       &G\ket{0}=\ket{G}=\sum_j \sqrt{a_i}\ket{j}_1\ket{\chi_j}_2,\\
       &\rho = \mathrm{Tr}(\ket{G}\bra{G})_1=\sum_j a_i\ket{\chi_j}_2\bra{\chi_j}_2.
\end{split}
\end{eqnarray*}
Let $S$ be a swap gate between register 2 and an ancillary system, i.e. register 3, then $(G^\dagger \otimes I_3)(I_1 \otimes S_{2,3})(G \otimes I_3)$ is a $(1,n+s,0)$ block-encoding of $\rho$.
\end{lemma}

\begin{lemma}(\cite{chakraborty2019})
\label{linearsystem}
Let $A$ be an $n\times n$ Hermitian matrix with non-zero eigenvalues lying in $[-1,-1/\kappa]\bigcup[1/\kappa,1]$, $\kappa \ge 2$. Assume that we have a unitary $U$ which is an $(\alpha, a, \delta)$ block-encoding of $A$ that can be implemented in time $O(T_U)$, where $\delta=O(\epsilon/(\kappa^2 \log_{2}^3 \frac{\kappa}{\epsilon}))$. Also, assume that we can prepare the state $\ket{b}$ which spans the eigenvectors with non-zero eigenvalues of $A$ in time $O(T_b)$. Then there is a quantum algorithm that output the quantum state $\frac{A^{-1}\ket{b}}{\|A^{-1}\ket{b}\|}$ with error $\epsilon$ in time
\begin{eqnarray}
\begin{split}
        O\left(\kappa (\alpha(T_u+a)\log_{2}^2 (\frac{\kappa}{\epsilon}) +  T_b)\log_{2} \kappa\right).
\end{split}
\end{eqnarray}
\end{lemma}

\begin{lemma}(\cite{Quantumspe,KPAQI})
\label{tomography}
Assume that there is a quantum algorithm to prepare the quantum state $\ket{\mathbf{x}}=\sum_{i=0}^{d-1} x_i\ket{i}$ in time $O(T)$, then there is a quantum algorithm allows us to output a classical vector $\mathbf{x}=(x_0,x_1,...,x_{d-1})^T$ that satisfies $\|\mathbf{x}-\ket{\mathbf{x}}\| \le \delta$ in time $O(\frac{Td\log_{2} d}{\delta^2})$ with probability at least $1-1/\mathrm{poly}(d)$.
\end{lemma}

We now detail the stage 1. We first perform the $U_B$ in Eq. (\ref{oracleA}) on the state $\ket{i}\ket{0}$ to get the state
\begin{eqnarray}
\label{AfUa}
        \ket{i} \frac{1}{\sqrt{k^{(i)}}}\sum_{\mathbf{x}_j\in Q_i}\ket{j}.
\end{eqnarray}
According to \emph{Lemma} \ref{Le:distance}, we can prepare the state
    \begin{eqnarray}
    \label{eq:xx}
       \ket{i} \frac{1}{\sqrt{k^{(i)}}} \sum_{\mathbf{x}_j\in Q_i}\ket{j} \ket{\|\mathbf{x}_i-\mathbf{x}_j\|^2}.
    \end{eqnarray}
Using controlled rotation, we have
    \begin{eqnarray}
    \begin{split}
    \label{aa}
       &\ket{i}  \frac{1}{\sqrt{k^{(i)}}} \sum_{\mathbf{x}_j\in Q_i}\ket{j} \ket{\|\mathbf{x}_i-\mathbf{x}_j\|^2}\\
       &\otimes (\frac{\|\mathbf{x}_i-\mathbf{x}_j\|}{r} \ket{1}
       + \sqrt{1-\frac{\|\mathbf{x}_i-\mathbf{x}_j\|^2}{r^2}} \ket{0}).
    \end{split}
    \end{eqnarray}

Then by uncomputing the third register and measuring the last qubit to get $\ket{1}$, we can obtain the state
    \begin{eqnarray}
    \begin{split}
    \label{eq:nomalized1}
       \ket{i}  \frac{1}{\sqrt{c^{(i)}}} \sum_{\mathbf{x}_j\in Q_i} \|\mathbf{x}_i-\mathbf{x}_j\|\ket{j},
    \end{split}
    \end{eqnarray}
where $c^{(i)}=\sum_{\mathbf{x}_j\in Q_i}  \|\mathbf{x}_i-\mathbf{x}_j\|^2 $ is the normalization factor.

Finally, according to \emph{Lemma} \ref{minus}, we can append $\ket{\mathbf{x}_i-\mathbf{x}_j}$ to the component marked by $\ket{i}\ket{j}$ of the state in Eq. (\ref{eq:nomalized1}) to get the state $\ket{\psi^{(i)}}$.

In stage 2, assume the unitary to prepare the state $\ket{\psi^{(i)}}$ is $G'$. Since $\mathrm{Tr}(\ket{\psi^{(i)}}\bra{\psi^{(i)}})_{1,3}=\rho_{C^{(i)}}$,  according to \emph{Lemma} \ref{blockencode}, we can obtain a $(1,2\log_{2}(m)+\log_{2}(n),0)$ block-encoding of $\rho_{C^{(i)}}$, that is, $( {G'}^\dagger \otimes I_4)(I_{1,3} \otimes S_{2,4})({G'}\otimes I_4)$, $S$ is a swap gate operated on register 2 and an ancillary register 4. According to \emph{Lemma }\ref{linearsystem}, we can obtain $\ket{W_i}= \frac{\rho_{C^{(i)}}^{-1}\ket{B_i}}{\|\rho_{C^{(i)}}^{-1}\ket{B_i}\|}$.

In stage 3, we perform quantum tomography on $\ket{W_i}$ to get $W_i$. Although $\ket{W_i}$ is of dimension $n$, Only $k^{(i)}$ components have non-zero amplitudes and the positions of the non-zero amplitudes are known. Thus we can perform a unitary to transform $\ket{W_i}$ to a state that only the first $k^{(i)}$ amplitudes are nonzero that can be regarded as a $\log_{2} (k^{(i)})$-dimension state $\ket{W'_i}$. According to the \emph{Lemma} \ref{tomography}, we can obtain a classical vector $W'_i$. It should be noted that $W'_i$ satisfies $\sum_j W'_{ij}=1$, thus renormalization is needed after quantum tomography. The renormalized vector $W'_i$ is actually what we want since it contains all the information of $W_i$.
\subsubsection{Complexity analysis}
\label{subsec2}

In stage 1, to prepare the state in Eq. (\ref{AfUa}), $U_B$ is invoked for one time, thus the complexity is $O[\polylog(mn)]$.
According to \emph{Lemma} \ref{Le:distance}, the complexity to prepare the state in Eq. (\ref{eq:xx}) is $T_1=O\left[\frac{(\max_i \|\mathbf{x}_i\|)^2 T \log_{2}(1/\delta)}{\epsilon_1}\right]$, where $\epsilon_1$ is the error of $\|\mathbf{x}_i-\mathbf{x}_j\|^2$.
The complexity of the controlled rotation can be neglected. Let $\epsilon_0=\min_{i,j} {\|\mathbf{x}_i-\mathbf{x}_j\|}$, we have $\|\mathbf{x}_i-\mathbf{x}_j\|/r \ge \epsilon_0/r$, where $r$ is the fixed radius which can be regarded as a constant. The probability to measure the last qubit of the state in Eq. (\ref{aa}) in computational basis to get an output $\ket{1}$ is
\begin{eqnarray}
        p(1)=\sum_{\mathbf{x}_j\in Q_i} \frac{\|\mathbf{x}_i-\mathbf{x}_j\|^2}{r^2 k^{(i)}}\ge \epsilon_0^2/r^2.
\end{eqnarray}
Using quantum amplitude amplification, $O(r/\epsilon_0)=O(1/\epsilon_0)$ times of repetition is enough to get the state in Eq. (\ref{eq:nomalized1}). The last step is to append $\ket{\mathbf{x}_i-\mathbf{x}_j}$ to the state in  Eq. (\ref{eq:nomalized1}) to obtain $\ket{\psi^{(i)}}$.
According to  \emph{Lemma }\ref{minus}, it takes time $O(\frac{h}{\epsilon_0}\polylog(mn/\epsilon))$, where $h=\max_{i} \|\mathbf{x}_i\|$. Let $\epsilon_1=O(\epsilon^2 \epsilon_0^2)$, $\delta=O(\epsilon_0 \epsilon)$, then the error of $\ket{\psi^{(i)}}$ is within $O(\epsilon)$.
As a conclusion, the complexity of stage 1, i.e., to prepare the state $
\ket{\psi^{(i)}}$, is $T^{(2)}_{s1}=O\left[\frac{h^2}{\epsilon^2 \epsilon_0^2} \polylog(\frac{mn}{\epsilon\epsilon_0}) \right]$.

For stage 2, since the error to prepare $\ket{\psi^{(i)}}$ is $O(\epsilon)$, according to \emph{Lemma \ref{blockencode}}, we have a $(1,2\log_{2} m + \log_{2} n,\epsilon)$ block-encoding of $\rho_C^{(i)}$. Note that $\rho_{C^{(i)}}$ is a sparse matrix that only $k^{(i)}$ rows and columns contains non-zero elements, thus the rank of $\rho_{C^{(i)}}$ is at most $k^{(i)}$. Without loss of generality, let the nonzero eigenvalues of $\rho_{C^{(i)}}$ be $\lambda_{0}\le\lambda_1\le...\le\lambda_{k^{(i)}-1}$, we have $\sum_{j=0}^{{k^{(i)}-1}} \lambda_j = 1$, $\max_i \lambda_i=\lambda_{k^{(i)}-1}\ge 1/k^{(i)}$. Assume the condition number of $C^{(i)}$ is $\kappa^{(i)}$, then $\lambda_j\in [\frac{1}{k^{(i)}\kappa^{(i)}}, 1]$. The state $\ket{B_i}$ can be prepared by oracle $U_B$ in time $O(\polylog (mn))$. Thus according to \emph{Lemma }\ref{linearsystem}, we can obtain $\ket{W_i}= \frac{\rho_{C^{(i)}}\ket{B_i}}{\|\rho_{C^{(i)}}\ket{B_i}\|}$ in time $T^{(2)}_2=O\left(\frac{h^2k^{(i)}\kappa^{(i)}}{\epsilon^2\epsilon_0^2} \polylog (\frac{mn}{\epsilon\epsilon_0}) \right).$

In stage 3, according to \emph{Lemma} \ref{tomography}, we can output the vector $W'_i$ in $T^{(2)}_3=O(T^{(2)}_2 k^{(i)}\log_{2} k^{(i)} /\epsilon^2)$ with probability at least $1-1/\mathrm{poly}(d)$, where $\epsilon$ is the error of vector $W'_i$. The complexity of renormalized can be neglected.

As a conclusion, the complexity to obtain the information of $W$ is $T^{(2)}=O(mT^{(2)}_3)=O\left(\frac{h^2m k_{\max}^2 \kappa_{\max}}{\epsilon^4 \epsilon_0^2}\polylog(\frac{mn}{\epsilon \epsilon_0})\right)$ where $k_{\max}=\max_i k^{(i)}$, $\kappa_{\max}=\max_i \kappa^{(i)}$.

\subsection{The quantum algorithm to compute the transformation matrix $A$}
\label{subsecC}

We have obtained the classical information of $W$ in the above algorithm. Thus we can store the information of the matrix $D=I-W$ in a data structure that allows the following two mappings:
\begin{eqnarray}
\begin{split}
\label{oracleM}
        &U_D: \ket{i}\ket{0} \mapsto  \ket{i}\ket{D_i},\\
        &V_D: \ket{0}\ket{j} \mapsto \frac{1}{\sqrt{\|D\|_F}}\sum_i \|D_i\|\ket{i}\ket{j}
\end{split}
\end{eqnarray}
in time $O[\polylog (mn)]$, where $\ket{D_i}$ is proportional to the $i$th row of $D$. Note that $W$ is a matrix of $K=\Theta(mk)$ nonzero elements and the diagonal elements are $0$, the space and time complexity to construct the data structure of $D$ are $O(mk\polylog(mk))$, the same as the matrix $B$.

Let $D=\sum_{j=0}^{m-1} \sigma_j \ket{\mathbf{u}_j}\bra{\mathbf{v}_j}$, where $0 \le \sigma_0 \le \sigma_1\le...\le\sigma_{m-1}$, then $M=D^T D=\sum_j \sigma_j^2 \ket{\mathbf{v}_j}\bra{\mathbf{v}_j}$.
According to \cite{ghojogh2020locally}, $D$ is a matrix with rank less than $m-1$. Without loss of generality, let the bottom $d$ nonzero eigenvalue of $M$ be $\sigma_1^2$ to $\sigma_{d}^2$ ($\sigma_0$ is 0) with corresponding eigenvectors $\ket{\mathbf{v}_1},...,\ket{\mathbf{v}_d}$.

To simplify the description of the quantum algorithm, we make a modification of the definition of the matrix $X$. If $m \le n$, we add $n-m$ zero rows to get an $n\times n$ matrix and vice versa. Here without loss of generality, we assume $m\le n$, then we get an $n\times n$ matrix $X$. We should mention that the data structure does not need to be modified.

Assume that we have obtained the $\ket{\mathbf{v}_1}$ to $\ket{\mathbf{v}_d}$ which are actually the  $\mathbf{z}_1,\mathbf{z}_2,...,\mathbf{z}_d$ (see Eq. (\ref{mz})). By appending $\log_{2} n-\log_{2} m$ zero states on $\ket{\mathbf{v}_i}$, we obtain a $\log_{2} n$-qubit state $\ket{\mathbf{v}_i^n}=\ket{0}^{\otimes (\log_{2} n-\log_{2} m)}\ket{\mathbf{v}_i}$.

Let $X=\sum_i \gamma_i \ket{\mathbf{u}^X_i}\bra{\mathbf{v}^X_i}$, $\ket{\mathbf{v}^n_j}=\sum_i \beta_i \ket{\mathbf{u}^X_i}$, then we have
\begin{eqnarray}
\begin{split}
        \mathbf{a}_j &=\left(X^T X +\alpha I\right)^{-1}X^T \ket{\mathbf{v}_j^n} \\
        &=\sum_i \frac{\beta_i \gamma_i}{\gamma_i^2+\alpha}\ket{\mathbf{v}^X_i}.
\end{split}
\end{eqnarray}
Let
\begin{eqnarray}
\label{hatX}
\begin{split}
       \overline{X}&:=
       \left[\begin{matrix}
       0 & X \\
       X^T & 0
       \end{matrix}\right]
       =
       \left[\begin{matrix}
       0 & \sum_i \gamma_i \ket{\mathbf{u}^X_i}\bra{\mathbf{v}^X_i} \\
       \sum_i \gamma_i \ket{\mathbf{v}^X_i}\bra{\mathbf{u}^X_i}\ & 0
       \end{matrix}\right]\\
       &=\sum_i \pm \gamma_i \ket{\psi_{i\pm}}\bra{\psi_{i\pm}},
\end{split}
\end{eqnarray}
where
\begin{eqnarray}
\begin{split}
\ket{\psi_{i\pm}}= \frac{1}{\sqrt{2}}\left[\begin{matrix}\ket{\mathbf{u}^X_i} \\ \pm\ket{\mathbf{v}^X_i}\end{matrix}\right]=\frac{1}{\sqrt{2}} (\ket{0,\mathbf{u}_i^X}\pm \ket{1,\mathbf{v}_i^X}).
\end{split}
\end{eqnarray}



\subsubsection{Algorithm details}

The algorithm can be decomposed to the following stages:
\begin{enumerate}
    \item Perform quantum singular value estimation (QSVE) \cite{WZP,IA} to get the state
        \begin{eqnarray}
        \begin{split}
               \ket{\psi_1}=\frac{1}{\sqrt{m}}\sum_{j=0}^{m-1} \ket{\mathbf{v}_j}\ket{\mathbf{v}_j} \ket{\sigma_j},
        \end{split}
        \end{eqnarray}
    \item Use the quantum algorithm for finding the minimum \cite{1996quant} to find the $d$ minimized nonzero value of $\sigma_j$, i.e., $\sigma_1,\sigma_2,...,\sigma_d$, and the corresponding $\ket{\mathbf{v}_j}$ for $j=1,2,...,d$.
    \item Perform quantum ridge regression \cite{YGW} to get $\ket{\mathbf{a}_j} \propto \mathbf{a}_j = \left(X^T X +\alpha I\right)^{-1}X^T \mathbf{v}_i$ for $j=1,2,...,d$.
\end{enumerate}

Here we list lemmas that will be used in our algorithm:
\begin{lemma}(QSVE \cite{WZP,IA})
\label{QSVE}
Assume that an $m \times n$ matrix $D=\sum_{j} \sigma_j \ket{\mathbf{u}_j}\bra{\mathbf{v}_j}$ is stored in a data structure that allows the two mappings showed in Eq. (\ref{oracleM}) in time $O(\polylog (mn))$. Let $\delta>0$ be the precision number. Then there is a quantum algorithm that transforms $\sum_j \beta_j \ket{\mathbf{v}_j}\ket{0}$ to $\sum_j \beta_j\ket{\mathbf{v}_j}\ket{\overline{\sigma_j}}$, where $\overline{\sigma_j}\in \sigma_j\pm\delta \|D\|_F$ with probability at least $1-1/\mathrm{poly}(n)$ in time $O(\frac{1}{\delta}\polylog(mn))$ for $j=0,1,...,m-1$.
\end{lemma}
\begin{lemma}(\cite{chakraborty2019})
\label{A^T}
Let $A$ be an $m\times n$ matrix stored in a data structure showed in Eq. (\ref{oracleM}), then there exists $U_R$ and $U_L$ that can be implemented in time $O(\polylog(mn)/\epsilon)$ such that $U_L^\dagger U_R$ is a $(\|A\|_F, \lceil\log_{2}(m+n)\rceil,\epsilon)$ block-encoding of
$ \overline{A}=\left[\begin{matrix}    0 & A\\    A^T & 0    \end{matrix}\right]$.
\end{lemma}
\begin{lemma}(\cite{chakraborty2019})
\label{Xsimulation}
Suppose that $U$ is an $(\alpha, a, \epsilon/|2t|)$ block-encoding of $H$, then we can implement $e^{iHt}$ with $O(|\alpha t|+\log_{2}(1/\epsilon))$ query of $U$ or its inverse and $O(a|\alpha t|+a\log_{2}(1/\epsilon))$ two-qubit gates, where $\epsilon$ is the error of $e^{iHt}$.
\end{lemma}
In stage 1, since we have no information about the state $\ket{\mathbf{v}_j}$ for $j=0,1,...,m-1$, we choose the state $\ket{\psi_0}=\frac{1}{\sqrt{m}}\sum_{j=0}^{m-1} \ket{j}\ket{j}$ to be the initial state, which can be written as uniform superposition of $\ket{\mathbf{v}_j}\ket{\mathbf{v}_j}$, i.e.,
\begin{eqnarray}
\ket{\psi_0}=\frac{1}{\sqrt{m}}\sum_{j=0}^{m-1} \ket{\mathbf{v}_j}\ket{\mathbf{v}_j}.
\end{eqnarray}
Note that we have stored the matrix $D$ in a suitable data structure and $D=\sum_{j=0}^{m-1} \sigma_j \ket{\mathbf{u}_j}\bra{\mathbf{v}_j}$, according to QSVE (\emph{Lemma} \ref{QSVE}), we can obtain the state $\ket{\psi_1}$.

For stage 2, let $U_2$ be the unitary to prepare $\ket{\psi_1}$ from $\ket{0}$ and $O_2$ be the oracle to transform $\ket{\psi_1}$ to
\begin{eqnarray}
        \frac{1}{\sqrt{m}}(\sum_{\sigma_j>v,\sigma_j=0} \ket{\mathbf{v}_j}\ket{\mathbf{v}_j}  \ket{\sigma_j}-\sum_{0<\sigma_j\le v} \ket{\mathbf{v}_j}\ket{\mathbf{v}_j}  \ket{\sigma_j}),
\end{eqnarray}
where $v$ is a constant. Given the $U_2$ and $O_2$, we can invoke the quantum algorithm for finding the minimum \cite{1996quant} to find the minimized nonzero value of $\sigma_j$, i.e., $\sigma_1$. Assume that we have obtained $i$ minimum values of $\sigma_j$ for $j\in\{1,2,...m-1\}$, we could make a small modification on $O_2$ to get an oracle $O_2'$ that
\begin{eqnarray*}
       O_2'=\frac{1}{\sqrt{m}}(\sum_{\sigma_j>v,\sigma_j\le \sigma_i} \ket{\mathbf{v}_j}\ket{\mathbf{v}_j}  \ket{\sigma_j}-\sum_{\sigma_i<\sigma_j\le v} \ket{\mathbf{v}_j}\ket{\mathbf{v}_j}  \ket{\sigma_j}).
\end{eqnarray*}
Then we can perform the quantum algorithm for finding the minimum to find the $i+1$ minimized nonzero value of $\sigma_j$. Thus $O(d)$ times of the quantum algorithm for finding the minimum is enough to obtain the $\sigma_1$ to $\sigma_d$. We should mention that when we get a $\sigma_j\in\{\sigma_1,\sigma_2,...,\sigma_d\}$, we also get two quantum state $\ket{\mathbf{v}_j}$.

For stage 3, since $X$ is stored in the data structure, according to \emph{Lemma} \ref{A^T} we can implement an $(\|X\|_F, \lceil\log_{2}(2n)\rceil,\epsilon_2)$ block-encoding of $\overline{X}$. According to \emph{Lemma} \ref{Xsimulation}, we can implement $e^{i\overline{X}t}$. Then we can perform quantum ridge regression \cite{YGW} to obtain $\ket{\mathbf{a}_j}$. The algorithm proceeds as following steps:

(1) Prepare the $\log_{2}(n)+1$ dimensional quantum state $\ket{0,\mathbf{v}_j^n}=\sum_i \beta_i \ket{0,\mathbf{u}^X_i}=\frac{1}{\sqrt{2}}\sum_i \beta_i (\ket{\psi_{i+}}+\ket{\psi_{i-}})$ by expanding the quantum state $\ket{\mathbf{v}_j}$.

(2) Perform quantum phase estimation on the state $\ket{0,\mathbf{v}_j^n}$ by simulating $e^{i\overline{X}t}$ to get the eigenvalues and eigenvectors of $\overline{X}$, i.e., obtain the state
\begin{eqnarray}
                 \frac{1}{\sqrt{2}}\sum_i \beta_i \ket{\psi_{i\pm}}\ket{\pm\gamma_i}.
\end{eqnarray}

(3) Perform controlled rotation and uncompute the phase estimation to get the state
\begin{eqnarray*}
                 \frac{1}{\sqrt{2}}\sum_i \beta_i \ket{\psi_{i\pm}}
                 (\frac{\pm C_1\gamma_i}{\gamma_i^2+\alpha}\ket{0}+
                 \sqrt{1-\frac{C_1^2\gamma_i^2}{(\gamma_i^2+\alpha)^2}}\ket{1}),
\end{eqnarray*}
where $C_1$ is a constant.

(4) Measure the last qubit to get $\ket{0}$, and project the first register onto the $\ket{\mathbf{v}_j^X}$ part (i.e., measure the first qubit of $\ket{\psi_{i\pm}}$ to get $\ket{1}$), we can obtain $\ket{\mathbf{a}_j}$,
\begin{eqnarray}
\begin{split}
        \ket{\mathbf{a}_j} = \frac{1}{C}\sum_i \frac{\beta_i \gamma_i}{\gamma_i^2+\alpha}\ket{\mathbf{v}^X_i},
\end{split}
\end{eqnarray}
where $C$ is the normalized factor.


\subsubsection{Complexity analysis}

In stage 1, the preparation of $\ket{\psi_0}$ is of time $O(\log_{2} m)$. Let the error of $\sigma_j$ to be $\epsilon$, according to \emph{Lemma} \ref{QSVE}, the complexity to get $\ket{\psi_1}$ is $O(\frac{\|D\|_F}{\epsilon} \polylog (mn))$.

In stage 2, the $O_2$ (or $O'_2$) and $U_2$ can be implemented in time $O(\polylog(m))$ and $O(\frac{\|D\|_F}{\epsilon} \polylog (mn))$, respectively. The quantum algorithm for finding the minimum would output the minimum value with probability larger than $1/2$ with query complexity $O(\sqrt{m})$. We should mention that one query includes two $U_2$ and one $O_2$ (or $O'_2$). Thus to get $\sigma_1$ to $\sigma_d$ and $\ket{\mathbf{v}_1}$ to $\ket{\mathbf{v}_d}$, $O(d)$ times of the algorithm for finding the minimum is enough. The total complexity is $O(\frac{d\sqrt{m}\|D\|_F}{\epsilon}\polylog (mn))$.

In stage 3, for the step 1, we can append several $\ket{0}$ to $\ket{\mathbf{v}_j}$ to get $\ket{0,\mathbf{v}_j^n}$. For the step 2, an $(\|X\|_F, \lceil\log_{2}(2n)\rceil,\epsilon_2)$ block-encoding of $\overline{X}$ can be implemented in time $O(\polylog(mn)/\epsilon_2)$. According to \emph{lemma} \ref{Xsimulation}, we can simulate $e^{i\overline{X}t}$ in time $O(\|X\|_F t \polylog(mn/\epsilon_2))$, where error  $\epsilon_3=2t\epsilon_2$. Let $\kappa$ denote the condition number of $X$, to ensure the error of the final state $\ket{\mathbf{a}_j}$ is within $\epsilon$, the maximum simulation time of the quantum phase estimation should be $t=O(\kappa/\epsilon)$ and $\epsilon_3=\epsilon/\log_{2}(\kappa/\epsilon)$.
Thus the complexity of the quantum phase estimation is $O(\frac{\|X\|_F \kappa}{\epsilon} \polylog(mn/\epsilon))$. The complexity of step 3 is the same as step 3. In step 4, we could choose $C_1=O(\max_i (\frac{\gamma_i}{\gamma_i^2+\alpha}))^{-1}$, thus have $\frac{C_1\gamma_i}{\gamma_i^2+\alpha}=O(1/\kappa)$ \cite{YGW}. $O(\kappa^2)$ repetitions are needed to get a $\ket{0}$ and it can be improved to $O(\kappa)$ repetitions by quantum amplitude amplification. The projection is success with probability $1/2$.

As a conclusion, the complexity to get $\ket{\mathbf{a}_1},\ket{\mathbf{a}_2},...,\ket{\mathbf{a}_d}$ is

\begin{eqnarray*}
\begin{split}
        &O(\kappa \frac{d\sqrt{m}\|D\|_F}{\epsilon}\polylog (mn) + \frac{d\|X\|_F \kappa^2}{\epsilon} \polylog(\frac{mn}{\epsilon}))\\
        =&O(d(\frac{\sqrt{m}\|D\|_F\kappa  + \|X\|_F \kappa^2}{\epsilon}) \polylog(\frac{mn}{\epsilon})).
\end{split}
\end{eqnarray*}
Assume that $\|W\|_{\max}=O(1)$. Since $W$ is sparse matrix and $D=I-W$, we have $\|D\|_F=O(\sqrt{mk})$. As for the $\|X\|_F$, since $h=\max_{i} \|\mathbf{x}_i\|$, $d\|X\|_F=O(\sqrt{hm})$. Thus the complexity of the algorithm is $O(\frac{d\sqrt{hk} m \kappa^2}{\epsilon}\polylog(\frac{mn}{\epsilon}))$.

\subsection{The total complexity and comparison}
\label{subsecD}

The procedure of the quantum NPE algorithm can be summarized as follows:

\begin{algorithm}[H]
\caption{The procedure of quantum NPE}
\label{Algorithm2}
\begin{algorithmic}[1]
\Require
The data matrix $X$ is stored in a data structure;
\Ensure
The quantum states $\ket{\mathbf{a}_1},\ket{\mathbf{a}_2},...,\ket{\mathbf{a}_d}$ which represent each row of matrix $A$;
\State Prepare $\frac{1}{m}\sum_{i,j=0}^{m-1} \ket{i}\ket{j} \ket{\sqrt{\frac{K}{m^2}}}$;
\State Prepare $\frac{1}{\sqrt{K}}\sum_{i=0}^{m-1} \ket{i}\sum_{\mathbf{x}_j \in Q_i}\ket{j}$;
\State Measure the output in computational basis for several times to obtain the index $j$ of the neighbors of $\mathbf{x}_i$ for $i=0,1,...,m-1$;
\State Construct oracle $U_B$ and $V_B$;
\State Prepare $\ket{\psi^{(i)}}$ to obtain $\rho_{C^{(i)}}$ ;
\State Prepare $\ket{W_i}=\rho_{C^{(i)}}^{-1}\ket{B_i}$ for $i=0,1,...,m-1$;
\State Perform quantum state tomography on $\ket{W_i}$ to get the information of $W_i$ for $i=0,1,...,m-1$;
\State Perform quantum singular value estimation to get $\ket{\psi_1}$;
\State Use the quantum algorithm for finding the minimum \cite{1996quant} to find $\sigma_j$ and $\ket{\mathbf{v}_j}$ for $j=1,2,...,d$.
\State Perform quantum ridge regression to get $\ket{\mathbf{a}_j}$ for $j=1,2,...,d$.\\
\Return $\ket{\mathbf{a}_1},\ket{\mathbf{a}_2},...,\ket{\mathbf{a}_d}$.
\end{algorithmic}
\end{algorithm}

The quantum algorithm can be divided into three sub-algorithms and the complexity of each sub-algorithm can be seen in Table \ref{tab:stc}. Putting it all together, the complexity of the quantum NPE algorithm is $O\left((h^2 m^{3/2} k^{1/2}+\frac{dh^2 m k_{\max}^2 \kappa_{\max} \kappa^2}{\epsilon^4 \epsilon_0^2})\polylog(\frac{mn}{\epsilon \epsilon_0})\right)$.
\begin{table}[!htb]
\caption{\label{tab:stc}
The time complexity of the three sub-algorithms of the quantum NPE.}
\begin{ruledtabular}
\begin{tabular}{cc}
Sub-algorithm$^\mathrm{a}$ &Time complexity \\ \hline
Algorithm 1 &$O\left(h^2 m^{3/2} k^{1/2} \polylog(\frac{mn}{\epsilon})\right)$ \\
Algorithm 2  &$O\left(\frac{h^2 m k_{\max}^2 \kappa_{\max}}{\epsilon^4\epsilon_0^2}\polylog(\frac{mn}{\epsilon\epsilon_0})\right)$\\
Algorithm 3& $O\left(\frac{d\sqrt{hk} m \kappa^2}{\epsilon}\polylog(\frac{mn}{\epsilon})\right)$ \\
\end{tabular}
\end{ruledtabular}
$^\mathrm{a}$Here the algorithm 1-3 are the quantum algorithm to find the nearest neighbors, the algorithm to obtain the weight matrix $W$ and the algorithm for embedding, respectively. $h=\max_{i} \|\mathbf{x}_i\|$, $k=\Theta(k^{(i)})$, $k^{(i)}$ is the number of neighbors of $\mathbf{x}_i$, $m$ is the number of training data points, $n$ is the dimension of the data points, $\epsilon$ is the error of the algorithm, $\epsilon_0=\min_{ij}\|\mathbf{x}_i-\mathbf{x}_j\|$, $k_{\max}=\max_i k^{(i)}$, $\kappa_{\max}=\max_i \kappa^{(i)}$, $\kappa^{(i)}$ is the condition number of the neighborhood
correlation matrix $C^{(i)}$, $d$ is the dimension of the low-dimensional space, $\kappa$ is the condition number of train data matrix $X$.
\end{table}

Since the classical algorithm have complexity $O(mnk^3 +dm^2)$, our algorithm have a polynomial speedup on $m$ and exponential speedup on $n$ when the factors $d,h,k_{\max},\kappa_{\max},\epsilon,\epsilon_0= O[\polylog(mn)]$. We should mention that the output of the quantum NPE algorithm is a matrix $A=(\ket{\mathbf{a}_1},\ket{\mathbf{a}_2},...,\ket{\mathbf{a}_d})$ with each column outputted as a quantum state.

Our algorithm has two advantages over VQNPE. (1) Our algorithm is complete while VQNPE is not. In \cite{Liang2020}, the authors pointed out that it is not known how to obtain the input of the third sub-algorithm from the output of the second sub-algorithm.
(2) The complexity of our algorithm is less than the complexity of VQNPE, even without considering the complexity of the third sub-algorithm of VQNPE. Specifically, The complexity of the first sub-algorithm is $O(\frac{m^2}{\epsilon^2}\log_{2} n)$, and the complexity of the second sub-algorithm is $\Omega(\mathrm{poly} (n))$ (we should mention that the complexity showed here are different with the original paper \cite{Liang2020}, see Appendix \ref{app:a} for details), while the total complexity of our algorithm is $O(m^{1.5}\polylog(mn))$ (only consider the main parameters). The advantage of our first sub-algorithm is mainly coming from the parallel estimation of the distance of each pair of data points. As for the second sub-algorithm, Liang et al. adopted the QSVD to get the $\ket{W_i}$. However, the eigenvalues of $\mathcal{A}_i$ (see Appendix \ref{app:a}) are too small to satisfy the conditions to get an efficient algorithm, which causes the complexity to have polynomial dependence on $n$. We use a totally different algorithm to get the $\ket{W_i}$ and the complexity analysis shows that our algorithm has complexity polylogarithmic dependence on $n$. As for the third sub-algorithm, it is hard to exam the complexity of the VQA of VQNPE, while our sub-algorithm has a rigorous complexity analysis.

\section{Conclusion}
In this paper, we proposed a complete quantum NPE algorithm with rigorous complexity analysis. It was showed that when $d,h,k_{\max},\kappa_{\max},\epsilon,\epsilon_0= O[\polylog(mn)]$, our algorithm has exponential acceleration on $n$ and polynomial acceleration on $m$ over the classical NPE. Also, our algorithm has a significant speedup compared with even the first two sub-algorithms of VQNPE.

The \emph{Lemma} \ref{minus} proposed an efficient method to append a quantum state generated by subtracting two vectors parallelly, which might have a wide range of applications in other quantum algorithms. Also, in the proof of the \emph{Lemma} \ref{minus}, we used a technique called parallel amplitude amplification, which may be of independent interest. We hope the techniques used in our algorithm could inspire more DR techniques to get a quantum advantage, especially the nonlinear DR techniques. We will explore the possibility in the
future.

\section*{Acknowledgements}
This work is supported by the Fundamental Research Funds for the Central Universities (Grant No.2019XDA01) and
National Natural Science Foundation of China (Grant Nos. 61972048, 61976024).
\appendix
\section{The proof of \emph{Lemma} \ref{minus}}
\label{app:aa}

\begin{proof}
$\ket{\mathbf{x}_i-\mathbf{y}_j}$ is a quantum state that is proportional to vector $\mathbf{x}_i-\mathbf{y}_j$,
   \begin{eqnarray}
   \begin{split}
            \ket{\mathbf{x}_i-\mathbf{y}_j} &=\ket{\|\mathbf{x}_i\| \ket{\mathbf{x}_i}- \|\mathbf{y}_i\| \ket{\mathbf{y}_i}}\\
            &=\frac{\|\mathbf{x}_i\| \ket{\mathbf{x}_i}- \|\mathbf{y}_i\| \ket{\mathbf{y}_i}}{\|\|\mathbf{x}_i\| \ket{\mathbf{x}_i}- \|\mathbf{y}_i\| \ket{\mathbf{y}_i}\|}.
   \end{split}
   \end{eqnarray}
\subsection{Algorithm details}
Let
   \begin{eqnarray}
        \ket{\psi}:=\frac{1}{\sqrt{m}}\sum_{i,j=0}^{m-1} \sqrt{p_{ij}}\ket{i}\ket{j}\ket{\mathbf{x}_i-\mathbf{y}_j},
   \end{eqnarray}
the process to prepare $\ket{\psi}$ from $\frac{1}{\sqrt{m}}\sum_{i,j=0}^{m-1} \sqrt{p_{ij}}\ket{i}\ket{j}$ can be summarized as follows:
\begin{enumerate}
    \item Given quantum state $\sum_{i,j=0}^{m-1} \sqrt{p_{ij}}\ket{i}\ket{j}$, prepare $\sum_{i,j=0}^{m-1} \sqrt{p_{ij}}\ket{i}\ket{j}\ket{\|\mathbf{x}_i\|} \ket{ \|\mathbf{y}_j\|}$.
    \item Prepare the following quantum state by controlled rotation \cite{HHL},
       \begin{eqnarray}
       \begin{split}
             \sum_{i,j=0}^{m-1} \sqrt{p_{ij}}\ket{i}\ket{j} \ket{\|\mathbf{x}_i\|} \ket{ \|\mathbf{y}_j\|}
              (\cos \theta_{ij} \ket{0}+ \sin \theta_{ij} \ket{1}),
       \end{split}
       \end{eqnarray}
        where $\cos \theta_{ij}= \frac{\|\mathbf{x}_i\|}{\sqrt{\|\mathbf{x}_i\|^2+\|\mathbf{y}_i\|^2}}$, and thus $\sin\theta_{ij}= \frac{\|\mathbf{y}_j\|}{\sqrt{\|\mathbf{x}_i\|^2+\|\mathbf{y}_j\|^2}} $.
   \item Uncompute the third and the fourth registers, and then query the oracles to obtain the state
          \begin{eqnarray}
             \sum_{i,j=0}^{m-1} \sqrt{p_{ij}}\ket{i}\ket{j} (\cos \theta_{ij} \ket{0} \ket{\mathbf{x}_i}+ \sin \theta_{ij} \ket{1} \ket{\mathbf{y}_j}).
         \end{eqnarray}
   \item Apply Hadamard gate to the third register to obtain
        \begin{eqnarray}
        \begin{split}
        \label{BefM}
             &\sum_{i,j=0}^{m-1} \sqrt{p_{ij}}\ket{i}\ket{j} \frac{1}{\sqrt{2}}\big[\ket{0}(\cos \theta_{ij} \ket{\mathbf{x}_i}+ \sin \theta_{ij} \ket{\mathbf{y}_j})\\
             &+\ket{1}(\cos \theta_{ij} \ket{\mathbf{x}_i}-\sin \theta_{ij} \ket{\mathbf{y}_j})\big]\\
             :=&\sum_{i,j=0}^{m-1} \sqrt{p_{ij}}\ket{i}\ket{j} (\cos \psi_{ij}\ket{\phi_{ij}^{+}}+\sin \psi_{ij}\ket{\phi_{ij}^{-}}).
        \end{split}
        \end{eqnarray}
        where $\cos \psi_{ij} \ket{\phi_{ij}^{+}}=\frac{1}{\sqrt{2}}\ket{0}(\cos \theta_{ij} \ket{\mathbf{x}_i}+\sin \theta_{ij} \ket{\mathbf{y}_j})$, $\sin \psi_{ij} \ket{\phi_{ij}^{-}}=\frac{1}{\sqrt{2}}\ket{1}(\cos \theta_{ij} \ket{\mathbf{x}_i}-\sin \theta_{ij} \ket{\mathbf{y}_j})$.
   \item Perform parallel quantum amplitude amplification to  get state
       \begin{eqnarray}
        \begin{split}
             &\sum_{i,j=0}^{m-1} \sqrt{p_{ij}}\ket{i}\ket{j} \ket{\phi_{ij}^{-}}\\
             =&\sum_{i,j=0}^{m-1} \sqrt{p_{ij}}\ket{i}\ket{j}\ket{1}\ket{\mathbf{x}_i-\mathbf{y}_j}.
        \end{split}
        \end{eqnarray}
   \item Discard the third register, the state left is $\sum_{i,j=0}^{m-1} \sqrt{p_{ij}}\ket{i}\ket{j}\ket{\mathbf{x}_i-\mathbf{y}_j}$.
\end{enumerate}

To make the step 5 (parallel quantum amplitude amplification) more clear, we give details here.
Let $U_1$ be the unitary that prepares the state in Eq. (\ref{BefM}) from quantum state $\sum_{i,j=0}^{m-1} \sqrt{p_{ij}}\ket{i}\ket{j}$, and $O$ be the unitary that transforms the state in Eq. (\ref{BefM}) to
       \begin{eqnarray}
        \begin{split}
             \sum_{i,j=0}^{m-1} \sqrt{p_{ij}}\ket{i}\ket{j} (\cos \psi_{ij}\ket{\phi_{ij}^{+}}-\sin \psi_{ij}\ket{\phi_{ij}^{-}}).
        \end{split}
        \end{eqnarray}
We first perform a parallel quantum amplitude estimation \cite{YGWW} to obtain the amplitudes of the target states. And then we perform the fixed-point quantum search \cite{PhysRevLett.113.210501} parallelly to obtain the final state.
We defined the Grover operator of the parallel quantum amplitude estimation as
       \begin{eqnarray}
        \begin{split}
        \label{grovero}
             U_1(I_{m^2\times m^2}\otimes(2\ket{0}\bra{0}^{\otimes (1+\log_{2} n)}-I_{2n\times 2n}))U_1^{\dagger}O.
        \end{split}
        \end{eqnarray}
The parallel quantum amplitude amplification consist of three steps:

1) Perform quantum amplitude estimation on the quantum state in Eq. (\ref{BefM}) to get the estimated values of $|\sin \psi_{ij}|:=\frac{1}{\sqrt{2}}\|\cos \theta_{ij} \ket{\mathbf{x}_i}-\sin \theta_{ij} \ket{\mathbf{y}_i})\|$ parallelly for each $\ket{i}\ket{j}$, i.e., obtain the state
        \begin{eqnarray*}
        \begin{split}
             \sum_{i,j=0}^{m-1} \sqrt{p_{ij}}\ket{i}\ket{j} (\cos \psi_{ij}\ket{\phi_{ij}^{+}}+\sin \psi_{ij}\ket{\phi_{ij}^{-}})\ket{|\sin \psi_{ij}|}.
        \end{split}
        \end{eqnarray*}

2) Let $L_{ij} = 2\lceil \frac{\log_{2}(2/\delta')}{|\sin \psi_{ij}|}\rceil$, prepare the state
\begin{eqnarray}
        \begin{split}
             \sum_{i,j=0}^{m-1} \sqrt{p_{ij}}\ket{i}\ket{j} (\cos \psi_{ij}\ket{\phi_{ij}^{+}}+\sin \psi_{ij}\ket{\phi_{ij}^{-}})\ket{L_{ij}},
        \end{split}
        \end{eqnarray}
where $\delta'>0$ is a parameter that related to the final error.

3) Controlled by $\ket{L_{ij}}$ and $\ket{i}\ket{j}$, we perform $S_{L_{ij}}=G(\alpha_l,\beta_l)...G(\alpha_1,\beta_1)=\prod_{k=1}^{l_{ij}} G(\alpha_k,\beta_k)$ on the third register of the above equation, where $l_{ij}=\lceil\frac{L_{ij}-1}{2}\rceil$, for all $k=1,2,...,l$,
\begin{eqnarray*}
        \begin{split}
             \alpha_k=-\beta_{l-k+1}=2\cot^{-1}\left(\tan(2\pi k/L_{ij})\sqrt{1-\gamma^2} \right),
        \end{split}
        \end{eqnarray*}
$\gamma^{-1}=T_{1/{L_{ij}}}(1/\delta')$, $T_L(x)= \cos(L\cos^{-1}(x))$ is the $L$th Chebyshev polynomial of the first kind. We can obtain
\begin{eqnarray*}
        \begin{split}
             \sum_{i,j=0}^{m-1} \sqrt{p_{ij}}\ket{i}\ket{j} \ket{\widetilde{\phi_{ij}^{-}}},
        \end{split}
        \end{eqnarray*}
where $\|\inn{\widetilde{\phi_{ij}^{-}}}{\phi_{ij}^{-}}\|^2\ge 1-{\delta'}^{2}$.

\subsection{complexity analysis}
In step 1, two times of queries are invoked. The complexity of step 2 can be neglected. In step 3, four queries are invoked. The Hadamard gate in step 4 is of complexity $O(1)$.

As for the step 5, we should amplify the amplitudes $\sin \psi_{ij}$ that satisfies
        \begin{eqnarray}
        \begin{split}
             |\sin \psi_{ij}|=&\|\frac{1}{\sqrt{2}}\ket{1}(\cos \theta_{ij} \ket{\mathbf{x}_i}-\sin \theta_{ij} \ket{\mathbf{y}_j})\|\\
             =&\|\frac{1}{\sqrt{2}}\frac{\mathbf{x}_i-\mathbf{y}_j}{\sqrt{\|\mathbf{x}_i\|^2+\|\mathbf{y}_j\|^2}})\|\\
             \ge & \frac{\epsilon_0}{2h},
        \end{split}
        \end{eqnarray}
where $\epsilon_0=\min_i \|\mathbf{x}_i-\mathbf{y}_i\|$, $h=\max_i \{\|\mathbf{x}_i\|,\|\mathbf{y}_i\|\}$. The complexity of one query of the Grover operator in Eq. (\ref{grovero}) is $O(\polylog(mn))$. According to \cite{PhysRevLett.113.210501}, we should ensure that $L_{ij}\ge \frac{\log_{2}(2/\delta')}{|\sin \psi_{ij}|}$. We could estimate $|\sin \psi_{ij}|$ within error $\frac{1}{2}|\sin \psi_{ij}|$, then choose $L_{ij}= 2\lceil\frac{\log_{2}(2/\delta')}{|\sin \psi_{ij}|}\rceil$ to ensure $L_{ij}\ge \frac{\log_{2}(2/\delta')}{|\sin \psi_{ij}|}$. With the error $\frac{1}{2}|\sin \psi_{ij}|$, the complexity of the parallel quantum amplitude estimation is $O(\frac{h}{\epsilon_0}\polylog(mn))$. The complexity of each query of $G(\alpha_k,\beta_k)$ is $\polylog(mn)$, thus the complexity of step 3) is
$O(\max_{ij} L_{ij}\polylog(mn))=O(\frac{h}{\epsilon_0}\log_{2}(1/\delta')\polylog(mn))$.

The complexity of step 6 can be neglected.

To ensure that the error of $\ket{\psi}$ is within $\epsilon$, we could just let $\delta'=O(\epsilon)$. As a conclusion, the complexity of the algorithm is $O(\frac{h}{\epsilon_0} \polylog(mn/\epsilon))$.
\end{proof}

\section{A brief complexity analysis of the first two sub-algorithms of VQNPE}
\label{app:a}

For the first sub-algorithm of VQNPE, i.e., the algorithm to find the $k$-nearest neighbors, the authors used swap test circuits to obtain the square of the inner product of each pair of data points, which was regarded as the distance between data points by the following steps. Here we should mention that the authors implicitly assumed that $\|\mathbf{x}_i\|=1$, since they defined $\ket{\mathbf{x}_i}=\sum_j x_{ij} \ket{j}$.  The complexity of this step should be $O(\frac{m^2}{\epsilon^2} \log_{2} n)$, where $\epsilon$ is the error of the square of inner products. Here we also point out that this method can not obtain the distance of the vectors such as $(1,0)^T$ and $(-1,0)^T$, thus we used a different method to get the distance without this problem.

Let $\mathbf{x}_j^{(i)}, j \in \{0,1,2,...,k^{(i)}-1\}$ denotes the $k^{(i)}$ nearest neighbors of $\mathbf{x}_i$, $\mathcal{A}_i=(\mathbf{x}_i-\mathbf{x}_0^{(i)},\mathbf{x}_i-\mathbf{x}_1^{(i)},...,\mathbf{x}_i-\mathbf{x}_{k^{(i)}-1}^{(i)}) $, $C_1^{(i)}=\mathcal{A}_i^\dagger \mathcal{A}_i$. Then $C_1^{(i)}$ is a $k^{(i)} \times k^{(i)}$ matrix which is actually the matrix left by deleting the zero rows and columns of $C^{(i)}$, i.e. deleting $j$ rows and $j$ columns  for $\mathbf{x}_j \notin Q_i$.

For the algorithm to obtain $\ket{\mathbf{w}_i}$, the authors assumed that there is an oracle to access the element of $\mathcal{A}_i$ for $i=0,1,...,m-1$, that is,
\begin{eqnarray}
\begin{split}
       \ket{j}\ket{l}\ket{0}\rightarrow \ket{j}\ket{l}\ket{\mathcal{A}^i_{jl}}=\ket{j}\ket{l}\ket{x_{ij}-{x^{(i)}_{lj}}},
\end{split}
\end{eqnarray}
where $\mathcal{A}^i_{jl}$ is the $j$th row $l$th column element of $\mathcal{A}_i$, $x_{ij}$ is the $j$th element of $\mathbf{x}_i$ and $x^{(i)}_{lj}$ is the $j$th element of the $l$th nearest neighbor of $\mathbf{x}_i$.

With the oracle mentioned above, according to quantum singular value decomposition (QSVD) \cite{RSML}, since $\mathcal{A}_i$ is an $n \times k$ matrix, one could simulate $e^{i{\frac{\hat{\mathcal{A}}_i}{n+k}t}}$ with complexity $O(\frac{t^2}{\epsilon} \|\hat{\mathcal{A}}_i\|_F)$, where
\begin{eqnarray}
       \hat{\mathcal{A}}_i= \left(
       \begin{array}{cc}
            \mathbf{0} & \mathcal{A}_i \\
            \mathcal{A}_i^\dagger & \mathbf{0}
        \end{array}
            \right).
\end{eqnarray}
Let $\hat{\mathcal{A}}_i=\sum_{j=1}^{k} \sigma_{j\pm} \ket{\psi_{j\pm}}\bra{\psi_{j\pm}}$, similar to Eq. (\ref{hatX}). According to \cite{RSML}, the necessary condition for this algorithm to be efficient is that $\sigma_{j\pm}=\Theta(n+k)$ for $j\in \{1,2,...,k\}$.

Let $\lambda_j$, $j\in \{1,2,...,n+k\}$ denote all of the eigenvalue of $\hat{\mathcal{A}}_i$, it is obvious that $\sigma_{j\pm}$ is included by the set of $\lambda_j$. According to the Gershgorin circle theorem \cite{gerschgorin31}, for $j \le n$,
\begin{eqnarray}
\begin{split}
       |\lambda_j| \le \sum_{l=1}^{k} |\mathcal{A}_{jl}| \le \sum_{l=1}^{k^{(i)}} |x_{ij}-x_{lj}^{(i)}|.
\end{split}
\end{eqnarray}
since $\|\mathbf{x}_i\|=1$, we have $|x_{ij}| \le 1$, thus $|\lambda_j|\le \sum_{l=1}^{k} 2=2k$. Similarly, for $j > n$,
\begin{eqnarray*}
\begin{split}
       |\lambda_j| \le \sum_{l=1}^n |\mathcal{A}_{(j-n)l}| =\sum_l |x_{il}-x_{jl}^{(i)}|=\|\mathbf{x}_i-\mathbf{x}_j^{(i)}\|_1.
\end{split}
\end{eqnarray*}
Since $\|\mathbf{x}_i\|=1$, we have $\|\mathbf{x}_i-\mathbf{x}_j^{(i)}\|=\sum_l (x_{il}-x_{jl}^{(i)})^2 \le 2$, where $\mathbf{x}_j^{(i)}$ is the $j$th nearest neighbor of $\mathbf{x}_i$. According to the inequality $\|\mathbf{x}\|_1 \le \sqrt{n}\|\mathbf{x}\|$, we have $|\lambda_j| \le 2\sqrt{n}$. Thus we have $|\lambda_j| \le 2\sqrt{n}$ for all $j\in \{1,2,...,n+k\}$, which means that $\sigma_{i\pm}=O(\sqrt{n})\neq \Theta(n+k)$.

Since $\sigma_{i\pm}\neq \Theta(n+k)$, the algorithm is not efficient, thus the complexity is of $\Omega(\mathrm{poly}(n))$.

\bibliography{refe}
\end{document}